\documentclass[11pt,twoside]{article}

\usepackage{asp2006}
\usepackage{graphicx}

\begin{document}

\title{First Direct Detection of Magnetic Fields in Starspots and Stellar Chromospheres}

%\author{S. V. Berdyugina$^{1,2}$, D. M. Fluri$^1$,  N. Afram$^1$, and F. Suwald$^1$} 
\author{S. V. Berdyugina, D. M. Fluri,  N. Afram, and F. Suwald} 
\affil{Institute of Astronomy, ETH Zurich, CH-8092 Zurich, Switzerland}
%\affil{$^2$ Tuorla Observatory, University of Turku, FI-21500 Piikki\"o, Finland}

\author{P. Petit and J. Arnaud} 
\affil{Laboratoire d'Astrophysique de Toulouse et Tarbes (LATT) OMP, F-31400 Toulouse, France}

\author{D. M. Harrington and J. R. Kuhn}
\affil{Institute for Astronomy, University of Hawaii, Honolulu, HI 96822 USA}

%%% Abstract to run on from here.
\begin{abstract}
Here we report on the first detection of circular polarization in 
molecular lines formed in cool magnetic regions (starspots) and in
chromospheric emission lines formed in hot plages on the surfaces 
of active stars.
%The observations were obtained with the high-resolution spectro-polarimeter 
%ESPADOnS recently installed at the Canada-France-Hawaii Telescope. 
Our survey of G-K-M stars included young main-sequence dwarfs and 
RS~CVn-type giants and subgiants. All stars were found to possess surface 
magnetic fields producing Stokes $V$ LSD signals in atomic lines of 0.05\%\ 
to 0.5\%. Several stars clearly showed circular polarization in molecular 
lines of 0.1\%\ to 1\%. The molecular Stokes $V$ signal is reminiscent of 
that observed in sunspots. Chromospheric magnetic fields were detected 
on most active targets in Stokes~$V$ profiles of emission lines with peak
polarization up to 2\%. The observed molecular circular polarization on 
M dwarfs indicates single-polarity magnetic fields covering at least 10\%\ 
of the stellar disk. Smaller signals on K stars imply that their magnetic 
fields are apparently weaker, more entangled than on M dwarfs, or 
more diluted by the bright photosphere.
\end{abstract}

%%% MAIN BODY OF TEXT 

\section{Introduction}

Studying magnetic activity on stars other than the Sun provides an 
opportunity for detailed tests of solar and stellar  dynamo models, 
since an extensive sample of stars of various activity levels provides 
a wider range of global stellar parameters. The multitude of magnetic
phenomena are observed on cool active stars, including starspots in the 
photosphere, chromospheric plages, coronal loops, UV, X-ray and radio 
emission, and flares.

\begin{table}
\centering
\begin{tabular}{lllr|lllr}
\hline
      Star        & Sp.\ class & $V/I_{\rm c}$&$V/I_{\rm c}$ &
      Star        & Sp.\ class & $V/I_{\rm c}$&$V/I_{\rm c}$\\
                  &            & atoms       & TiO &
                  &            & atoms       & TiO \\     
\hline
      EK Dra        & G1 V      & 0.09$^*$  &   $\le$0.2  & AU Mic	    & M1 V      & 0.39$^*$  &   0.4$^*$\\
      V478 Lyr      & G8 V      & 0.12$^*$  &   $\le$0.1  & FK Aqr	    & M2/M3Ve  & 0.30$^*$  &   0.5$^*$\\
      $\xi$ Boo A   & G8 V      & 0.06      &   $\le$0.1  & EV Lac	    & M3.5Ve   & 0.28$^*$  &   1.1$^*$\\
      $\xi$ Boo B   & K4 V      & 0.06      &   $\le$0.2  & V1054 Oph       & M3.5Ve   & 0.14$^*$  &   0.4$^*$\\
      61 Cyg A      & K5 V      & 0.03$^*$  &   $\le$0.1  & $\lambda$ And   & G8 IV    & 0.09$^*$  &   $\le$0.1\\
      61 Cyg B      & K7 V      & $\le$0.01 &   $\le$0.1  & HK Lac	    & K0 III   & 0.08$^*$  &   $\le$0.1\\
      V833 Tau      & K5 V      & 0.29$^*$  &    0.2$^*$  & XX Tri	    & K0 III   & 0.14$^*$  &   $\le$0.2\\
      EQ Vir	    & K5 V      & 0.36$^*$  &   $\le$0.3  & 29 Dra	    & K1 III   & 0.14$^*$  &   $\le$0.1\\
      AX Mic        & K7 V      & 0.02$^*$  &   $\le$0.1  & BM CVn	    & K1 III   & 0.19$^*$  &   $\le$0.1\\
      BY Dra	    & K4/M0V    & 0.05$^*$  &   $\le$0.2  & IM Peg	    & K1 III   & 0.05      &   $\le$0.1\\
      SZ UMa	    & M0eV      & 0.04$^*$  &   $\le$0.2  & V1762 Cyg       & K1 IV    & 0.07$^*$  &   $\le$0.1\\
		    &           &	    &             & II Peg	    & K2 IV    & 0.36      &   0.1$^*$\\ 
\hline
\end{tabular}
\caption{Observed targets. Peak circular polarization $V/I_{\rm c}$ (\%) 
in atomic and molecular lines was measured from LSD profiles 
and the TiO 7055\,\AA\ band, respectively. Upper limits are estimated from
the noise level. Asterisks mark cases with first detections.}
\label{tab:obs}
\end{table}

Starspots and chromospheric plages are the best studied proxies of 
stellar magnetism. Large stellar brightness variations and indirect 
imaging of stellar surfaces with the Doppler Imaging technique 
indicate immense starspot regions as compared to sunspot sizes 
\citep{Berdyugina2005}. 
Molecular lines provide additional evidence of cool spots 
on the surfaces of active stars. If the effective temperature of the stellar
photosphere is high enough, molecular lines can only be formed in
cool starspots. The first detection of molecular bands from starspots was 
reported by \citet{Vogt1979} for a K2 star whose spectral type was not
compatible with the presence of TiO and VO bands indicating an equivalent 
spectral type of the spot spectrum as late as M6.
%From the relative strengths 
%and overall appearance of the molecular features an equivalent spectral type 
%of the spot spectrum was estimated as late as M6.

Spectropolarimetry in molecular lines which are only formed in starspot
umbrae can provide measurements of magnetic fields directly in spatially 
unresolved spots. For example, the strongest TiO band at 7055\,\AA\ is 
magnetically quite sensitive, having effective Land{\'e} factors up to 1 
\citep{BerdyuginaSolanki2002, Berdyuginaetal2003}. Thus, a clear Stokes $V$ 
signal in the TiO band is expected from starspots \citep{Berdyugina2002, 
Aframetal2006}. 

Stellar chromospheres are readily detected in emisison lines. 
Long-term monitoring of the Ca {\small II} H\&K emission revealed stellar activity
cycles on solar-like stars similar to the 11-year solar cycle 
\citep{Baliunas1995}. However, there is not yet evidence for global 
magnetic field reversals on stars similar to the Hale 22-year cycle
and no systematic measurements of stellar chromospheric fields.

Here we report the results of our spectropolarimetric survey of active 
G-K-M stars and the first detections of circular polarization in molecular and
chromospheric lines. 

\section{Observations}

Observations were carried out on July 14--16, 2005, and on August 1--3, 2006, 
at the Canada-France-Hawaii Telescope (CFHT) with the new spectropolarimeter ESPaDOnS.
Measurements were made in the circular polarization mode with four subsequent 
exposures at different waveplate angles. The calibration and reduction were made
with the 'libre esprit' software provided at the CFHT and included corrections for the 
dark current, flat-field, Fabry-P\'erot calibration, etc. The maximum polarimetric 
accuracy achieved was 10$^{-3}$.

Our survey included a sample of cool active stars: 15 G--M dwarfs and 8 G--K 
components of RS~CVn-type systems (Table~\ref{tab:obs}). The selected stars are 
moderate rotators ($v\sin i$\,$\le$\,24\,km/s) and brighter than $\sim$10th 
magnitude. All are known to have cool spots on their surfaces.

\begin{figure}
\begin{centering}
\includegraphics[width=13cm]{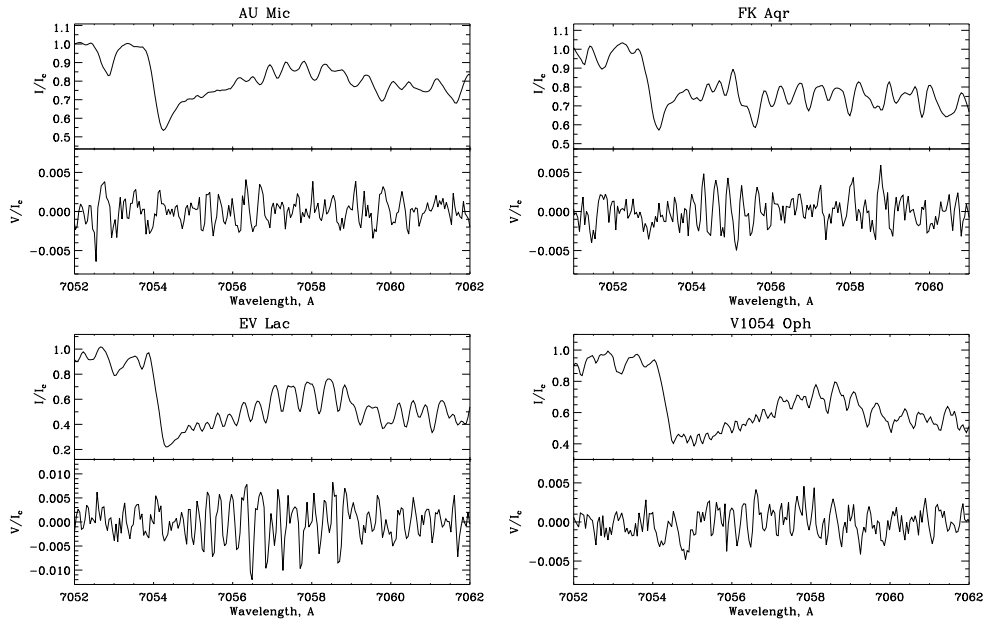}
\caption{Observed Stokes $I/I_{\rm c}$ (upper panels) and Stokes $V/I_{\rm c}$ 
(lower panels) of the TiO $\gamma$\,(0,0) $R_3$ band head on the active M dwarfs.
\label{fig:tio}}
\end{centering}
\end{figure}

\section{Results}

A clear Stokes $V$ signal in the TiO 7055\AA\ band (up to 1\%) was 
detected on four M dwarfs (Fig.~\ref{fig:tio}). Two stars (AU~Mic and EV~Lac) 
were known to have strong ($\sim$4\,kG) surface magnetic fields measured from 
Zeeman-broadened atomic lines \citep{Saar1992, JohnsKrullValenti1996} and these 
observations confirm those measurements. 
This is, however, the first detection of magnetic fields on both components of 
the FK~Aqr and V1054~Oph binaries, each consisting of two M dwarfs.
The shape of the Stokes $V$ signal is reminiscent of that observed in sunspots
\citep{Berdyuginaetal2000}.
A simple modeling of the observed circular polarization indicates single-polarity 
magnetic fields covering at least 10\% of the stellar disk \citep{Aframetal2006}.

A Stokes $V$ signal in the TiO band was also detected on two very active K stars 
from our sample, V833~Tau and II~Peg, with an amplitude of only 0.1--0.2\%\ 
(Table~\ref{tab:obs}). The detection on other stars was limited by the noise level
of 0.1--0.2\%\ on average. This implies that the magnetic fields on these stars 
are apparently weaker, more entangled than on M dwarfs, or more diluted by 
the bright photosphere.

\begin{figure}
\begin{centering}
\includegraphics[width=13cm]{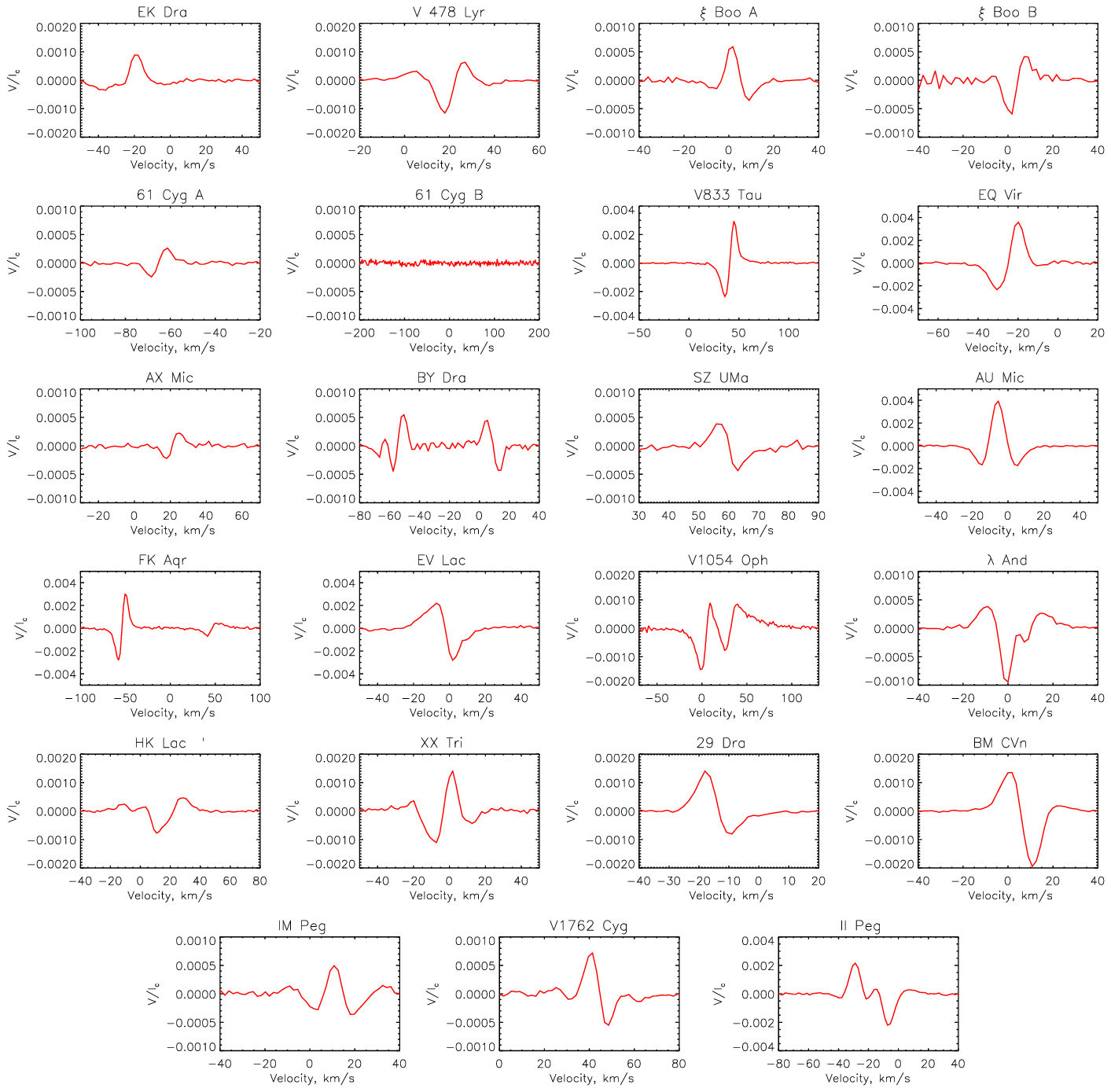}
\caption{Observed atomic LSD Stokes $V/I_{\rm c}$ profiles. Note detection of
magnetic fields on both components of the BY Dra, FK Aqr, and V1054~Oph binaries. 
\label{fig:lsd}}
\end{centering}
\end{figure}

A magnetic field was actually detected on all stars but one (61 Cyg B) in average 
atomic Stokes $V$ profiles extracted with the Least Squares Deconvolution (LSD) 
technique \citep{Donatietal1997}. For most stars this is the first detection 
of B-field from atomic lines
(Table~\ref{tab:obs}, Fig.~\ref{fig:lsd}). Note that the largest signals in the atomic
Stokes $V$ profiles were observed on those stars, where the TiO Stokes $V$
signals were prominent as well. On all these stars Stokes $V$
profiles in individual atomic lines were also recorded. A simultaneous analysis of 
the Stokes $I$ and $V$ signals from many atomic and molecular lines with 
different temperature and magnetic sensitivities will allow us to disentangle the 
contributions from the photosphere, faculae, and starspot umbrae and penumbrae.

\begin{figure}
\begin{centering}
\includegraphics[width=13cm]{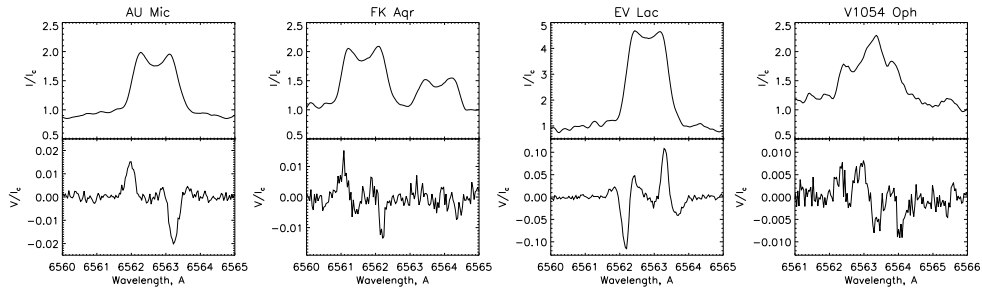}
\caption{Observed Stokes $I/I_{\rm c}$ (upper panels) and Stokes $V/I_{\rm c}$ 
(lower panels) of the H$\alpha$ line on the active M dwarfs.
\label{fig:ha}}
\end{centering}
\end{figure}

\begin{figure}
\begin{centering}
\includegraphics[width=13cm]{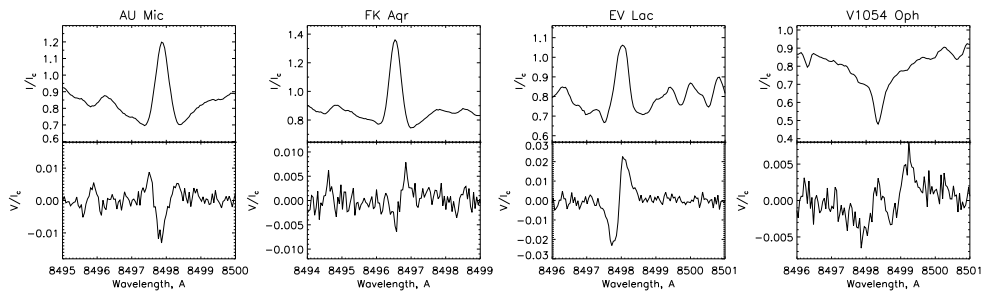}
\caption{Observed Stokes $I/I_{\rm c}$ (upper panels) and Stokes $V/I_{\rm c}$ 
(lower panels) in one of the infrared Ca {\small II} lines on the active M dwarfs.
\label{fig:ca}}
\end{centering}
\end{figure}

The stars from our sample are known to have active chromospheres as evidenced by 
emission in lines such as Ca {\small II} H\&K, H$\alpha$, infrared triplet 
Ca {\small II} lines, Na {\small I} D lines, etc. The most active stars, for which 
the largest LSD and TiO Stokes $V$ signals were detected, also exhibited strong 
polarization signals (up to 2\%) in these emission lines. Examples for the four 
active M dwarfs are shown in Figs.~\ref{fig:ha} and \ref{fig:ca}. An analysis 
of such data will provide direct mesurements of stellar chromospheric magnetic 
fields.

%%% Text of acknowledgements runs on after this command.
\acknowledgements 
We are thankful to the CFHT staff and our support astronomer Nadine Manset.
The observations were supported by the OPTICON access programme. This work was 
funded by ETH Research Grant TH-2/04-3 and SNF grants PE002-104552 and 200021-103696. 
S.V.~Berdyugina acknowledges the EURYI award from the ESF.

%%% THE BIBLIOGRAPHY


\begin{thebibliography}{}
\bibitem[Afram et al.(2006)]{Aframetal2006}
Afram, N., Berdyugina, S. V., Fluri, D. M., Solanki, S. K., Lagg, A., Petit, P., \& Arnaud, J.
   2006, in Solar Polarization Workshop 4, eds.\ R.\ Casini \& B.\ Lites, ASP.\ Conf.\ Ser., 
   358, 375 
\bibitem[Baliunas et al.(1995)]{Baliunas1995}
Baliunas, S. L., Donahue, R. A., Soon, W. H., et al. 1995, \apj, 438, 269
\bibitem[Berdyugina (2002)]{Berdyugina2002} 
Berdyugina, S. V. 2002, AN, 323, 192
\bibitem[Berdyugina(2005)]{Berdyugina2005} 
Berdyugina, S.V. 2005, Living. Rev. Solar Phys., 2, No 8
\bibitem[Berdyugina \& Solanki(2002)]{BerdyuginaSolanki2002}
Berdyugina, S. V., \& Solanki, S. K. 2002, A\&A, 385, 701
\bibitem[Berdyugina et al.(2000)]{Berdyuginaetal2000} 
Berdyugina, S. V., Frutiger, C., Solanki, S. K., \& Livingston, W. 2000, A\&A, 364, L101
\bibitem[Berdyugina et al.(2003)]{Berdyuginaetal2003} 
Berdyugina, S. V., Frutiger, C., \& Solanki, S. K. 2003, A\&A, 412, 513
%\bibitem[Berdyugina et al.(2005)]{berd05}
%Berdyugina, S.V., Braun, P.A., Fluri, D.M., \& Solanki, S.K. 2005, A\&A, 444, 947
%\bibitem[Donati(2006)]{Donati2006}
%Donati, J.-F. 2006, these proceedings
%\bibitem[Donati \& Collier Cameron(1997)]{DonatiCameron1997}
%Donati, J.-F., \& Collier Cameron, A. 1997, MNRAS, 291, 1
\bibitem[Donati et al.(1997)]{Donatietal1997}
Donati, J.-F., Semel, M., Carter, B. D., Rees, D. E., \& Collier Cameron, A. 1997, MNRAS, 291, 658
\bibitem[Johns-Krull \& Valenti(1996)]{JohnsKrullValenti1996}
Johns-Krull, C. M., \& Valenti, J. A. 1996, ApJ, 459, L95
\bibitem[Saar(1992)]{Saar1992} 
Saar, S. H. 1992, in Infrared solar physics, eds.\ D. M.\ Rabin, J. T.\ Jefferies, 
   \& C.\ Lindsey, IAU Symp.\ 154, 493
\bibitem[Vogt(1979)]{Vogt1979}
Vogt, S. S. 1979, PASP, 91, 616
\end{thebibliography}
\end{document}